\documentclass[a4paper,12pt, epsfig]{article}
\usepackage{epsfig}
\usepackage{amssymb,latexsym}
\usepackage{amsfonts}
\usepackage{amsmath}
\usepackage[colorlinks, linkcolor=blue, citecolor=blue, urlcolor=blue]{hyperref}
\usepackage{epstopdf}
\usepackage{graphicx}
\newskip\humongous \humongous=0pt plus 1000pt minus 1000pt

\newif\ifdtup

\catcode`\@=11

\@addtoreset{equation}{section}
\def\theequation{\arabic{section}.\arabic{equation}}

\def\@normalsize{\@setsize\normalsize{15pt}\xiipt\@xiipt
\abovedisplayskip 14pt plus3pt minus3pt%
\belowdisplayskip \abovedisplayskip
\abovedisplayshortskip \z@ plus3pt%
\belowdisplayshortskip 7pt plus3.5pt minus0pt}

\def\small{\@setsize\small{13.6pt}\xipt\@xipt
\abovedisplayskip 13pt plus3pt minus3pt%
\belowdisplayskip \abovedisplayskip
\abovedisplayshortskip \z@ plus3pt%
\belowdisplayshortskip 7pt plus3.5pt minus0pt
\def\@listi{\parsep 4.5pt plus 2pt minus 1pt
      \itemsep \parsep
      \topsep 9pt plus 3pt minus 3pt}}

\relax

\catcode`@=12

\catcode`\@=11

\def\section{\@startsection{section}{1}{\z@}{3.5ex plus 1ex minus
    .2ex}{2.3ex plus .2ex}{\large\bf}}

\def\thesection{\arabic{section}}
\def\thesubsection{\arabic{section}.\arabic{subsection}}

\def\appendix{\setcounter{section}{0}
  \def\thesection{Appendix \Alph{section}}
  \def\thesubsection{\Alph{section}.\arabic{subsection}}
  \def\theequation{\Alph{section}.\arabic{equation}}}
\def\SymBoxes#1#2#3#4{\newdimen\un@t \un@t#3%
\raisebox{#1}{\rule{#2\un@t}{#4}\hskip-#2\un@t% lower horizontal
\@tempdimb\un@t \advance\@tempdimb by-#4\@tempcntb#2\relax%
\@whilenum{\@tempcntb>0}\do{%                         % #2 vertical lines
\rule{#4}{\un@t}\hskip\@tempdimb \advance\@tempcntb by\m@ne}%
\hskip-#2\un@t \rule[\un@t]{#2\un@t}{#4}%
\rule[\un@t]{#4}{#4}\hskip-#4%             % upper horizontal line
\rule{#4}{\un@t}}\hskip-#4}                % rightest vertical line
\begin{document}
%\begin{letter}{~}

%%%%%%Define some new commands and  macros

\newcommand{\dd}{\textrm{d}}

\newcommand{\beq}{\begin{equation}}
\newcommand{\eeq}{\end{equation}}
\newcommand{\bea}{\begin{eqnarray}}
\newcommand{\eea}{\end{eqnarray}}
\newcommand{\beas}{\begin{eqnarray*}}
\newcommand{\eeas}{\end{eqnarray*}}
\newcommand{\defi}{\stackrel{\rm def}{=}}
\newcommand{\non}{\nonumber}
\newcommand{\bquo}{\begin{quote}}
\newcommand{\enqu}{\end{quote}}
\newcommand{\tc}[1]{\textcolor{blue}{#1}}
%%%%%%%%%%%%%%%%
\renewcommand{\(}{\begin{equation}}
\renewcommand{\)}{\end{equation}}
%%%%%%%%%%%%%%%%%%%%%%%%%%%%%%%%%% definitions
\def\errore{{\bf{Calcolami!}}}
\def\de{\partial}
\def\Om{\ensuremath{\Omega}}
\def\Tr{ \hbox{\rm Tr}}
\def\rc{ \hbox{$r_{\rm c}$}}
\def\H{ \hbox{\rm H}}
\def\HE{ \hbox{$\rm H^{even}$}}
\def\HO{ \hbox{$\rm H^{odd}$}}
\def\HEO{ \hbox{$\rm H^{even/odd}$}}
\def\HOE{ \hbox{$\rm H^{odd/even}$}}
\def\HHO{ \hbox{$\rm H_H^{odd}$}}
\def\HHEO{ \hbox{$\rm H_H^{even/odd}$}}
\def\HHOE{ \hbox{$\rm H_H^{odd/even}$}}
\def\K{ \hbox{\rm K}}
\def\Im{ \hbox{\rm Im}}
\def\Ker{ \hbox{\rm Ker}}
\def\const{\hbox {\rm const.}}
\def\o{\over}
\def\im{\hbox{\rm Im}}
\def\re{\hbox{\rm Re}}
\def\bra{\langle}\def\ket{\rangle}
\def\Arg{\hbox {\rm Arg}}
\def\exo{\hbox {\rm exp}}
\def\diag{\hbox{\rm diag}}
\def\longvert{{\rule[-2mm]{0.1mm}{7mm}}\,}
\def\a{\alpha}
\def\b{\beta}
\def\e{\epsilon}
\def\l{\lambda}
\def\ol{{\overline{\lambda}}}
\def\ochi{{\overline{\chi}}}
\def\th{\theta}
\def\s{\sigma}
\def\oth{\overline{\theta}}
\def\ad{{\dot{\alpha}}}
\def\bd{{\dot{\beta}}}
\def\oD{\overline{D}}
\def\opsi{\overline{\psi}}
\def\dag{{}^{\dagger}}
\def\tq{{\widetilde q}}
\def\L{{\mathcal{L}}}
\def\p{{}^{\prime}}
\def\W{W}
\def\N{{\cal N}}
\def\hsp{,\hspace{.7cm}}
\def\hspp{,\hspace{.5cm}}
\def\bo{\ensuremath{\hat{b}_1}}
\def\bfo{\ensuremath{\hat{b}_4}}
\def\co{\ensuremath{\hat{c}_1}}
\def\cfo{\ensuremath{\hat{c}_4}}
\def\th#1#2{\ensuremath{\theta_{#1#2}}}
\def\c#1#2{\hbox{\rm cos}(\th#1#2)}
\def\s#1#2{\hbox{\rm sin}(\th#1#2)}
\def\cp#1#2#3{\hbox{\rm cos}^#1(\th#2#3)}
\def\sp#1#2#3{\hbox{\rm sin}^#1(\th#2#3)}
\def\ctp#1#2#3{\hbox{\rm cot}^#1(\th#2#3)}
\def\cpp#1#2#3#4{\hbox{\rm cos}^#1(#2\th#3#4)}
\def\spp#1#2#3#4{\hbox{\rm sin}^#1(#2\th#3#4)}
\def\t#1#2{\hbox{\rm tan}(\th#1#2)}
\def\tp#1#2#3{\hbox{\rm tan}^#1(\th#2#3)}
\def\m#1#2{\ensuremath{\Delta M_{#1#2}^2}}
\def\mn#1#2{\ensuremath{|\Delta M_{#1#2}^2}|}
\def\u#1#2{\ensuremath{{}^{2#1#2}\mathrm{U}}}
\def\pu#1#2{\ensuremath{{}^{2#1#2}\mathrm{Pu}}}
\def\meff{\ensuremath{\Delta M^2_{\rm{eff}}}}
\def\an{\ensuremath{\alpha_n}}
\newcommand{\Z}{\ensuremath{\mathbb Z}}
\newcommand{\R}{\ensuremath{\mathbb R}}
\newcommand{\rp}{\ensuremath{\mathbb {RP}}}
\newcommand{\vac}{\ensuremath{|0\rangle}}
\newcommand{\vact}{\ensuremath{|00\rangle}                    }
\newcommand{\oc}{\ensuremath{\overline{c}}}
\renewcommand{\cos}{\textrm{cos}}
\renewcommand{\sec}{\textrm{sec}}
\renewcommand{\sin}{\textrm{sin}}
\renewcommand{\cot}{\textrm{cot}}
\renewcommand{\tan}{\textrm{tan}}
\renewcommand{\ln}{\textrm{ln}}

\renewcommand{\re}{\ensuremath{\mathcal{E}}}

\newcommand{\Vol}{\textrm{Vol}}

\newcommand{\half}{\frac{1}{2}}

%%%%%%%%%%%%%%%%%%%%%%%Changed%%%%%%%%%%%%%%%%%%%%%%%%%%%%%
\def\changed#1{{\bf #1}}
%\def\changed#1{ #1}
%%%%%%%%%%%%%%%%%%%%%%%%%%%%%%%%%%%%%%%%%%%%%%%%%%%%%%%%%

\begin{titlepage}
%\begin{flushright}
%IFUP-TH/2011-??
%\end{flushright}
%\bigskip

\def\thefootnote{\fnsymbol{footnote}}

\begin{center}
{\large {\bf
Measuring $\theta_{12}$ Despite an Uncertain Reactor Neutrino Spectrum
  } }
%\end{center}

\bigskip

\bigskip

{\large \noindent Emilio Ciuffoli$^{1}$\footnote{emilio@impcas.ac.cn}
,
 Jarah
Evslin$^{1}$\footnote{\texttt{jarah@impcas.ac.cn}},
Marco Grassi$^{2,3}$\footnote{\texttt{mgrassi@ihep.ac.cn}}
 and Xinmin Zhang$^{4,5}$\footnote{\texttt{xmzhang@ihep.ac.cn}} }
\end{center}

\renewcommand{\thefootnote}{\arabic{footnote}}

\vskip.7cm

\begin{center}
\vspace{0em} {\em  1)  Institute of Modern Physics, CAS, NanChangLu 509, Lanzhou 730000, China\\
2) Institute of High Energy Physics (IHEP), CAS, Beijing 100049, China\\
3) INFN Sezione di Milano, via Celoria 16, 20133 Milan, Italy\\
4) Theoretical physics division, IHEP, CAS,  Beijing 100049, China\\ 
5) Theoretical Physics Center for Science Facilities, IHEP, CAS,
    % P.O. Box 918(4), 
Beijing 100049, China\\

 {}}

%\vspace{0em} {\em  { 1) TPCSF, IHEP, Chinese Acad. of Sciences\\
%2) Theoretical physics division, IHEP, Chinese Acad. of Sciences\\
%YuQuan Lu 19(B), Beijing 100049, China}}

\vskip .4cm

\vskip .4cm

\end{center}

\vspace{1.3cm}

\noindent
\begin{center} {\bf Abstract} \end{center}

\noindent
The recently discovered 5 MeV bump highlights that the uncertainty in the reactor neutrino spectrum is far greater than some theoretical estimates.  Medium baseline reactor neutrino experiments will deliver by far the most precise ever measurements of $\theta_{12}$. However, as a result of the bump, such a determination of $\theta_{12}$ using the theoretical specrum would yield a value of $\spp2212$ which is more than 1\% higher than the true value.  We show that by using recent measurements of the reactor neutrino spectrum the precision of a measurement of $\theta_{12}$ at a medium baseline reactor neutrino experiment can be improved appreciably.  We estimate this precision as a function of the ${}^9$Li spallation background veto efficiency and dead time.

\vfill

\begin{flushleft}
{\today}
%\vspace{1cm}
\end{flushleft}
\end{titlepage}
%\bigskip

\hfill{}
%\bigskip

%\tableofcontents

\setcounter{footnote}{0}

\section{Introduction}
In about 5 years the largest liquid scintillator detectors ever built will be used to detect reactor neutrinos at the experiments JUNO \cite{juno} and RENO 50 \cite{reno50}.  The often-stated goal of these experiments is the determination of the neutrino mass hierarchy, following the strategy of Petcov and Piai \cite{petcovidea}.  Obtaining the required precision for a determination of the hierarchy will be very challenging \cite{parkedifficile,qianprimo,noisim}.  On the other hand, whether or not this precision can be achieved, there is no doubt that such experiments can provide by far the most precise measurement yet of $\theta_{12}$ \cite{idea12}.

In this note we will show that imperfect knowledge of the reactor neutrino spectrum is a leading source of uncertainty in the measurement of $\theta_{12}$ and that this uncertainty has been systematically underestimated in the literature.  Studies of this measurement use the latest reactor neutrino flux model from Ref.~\cite{huber}.  They also use the uncertainties quoted in that paper.  Nonetheless, as the author clearly stated in Ref.~\cite{hubertalk}, the uncertainty quoted in Ref.~\cite{huber} reflects only a subset of the sources of uncertainty in the analysis and so in fact yields only a lower bound on the true uncertainty.  Indeed, a widely accepted explanation for the reactor anomaly of Ref.~\cite{mention} is that the uncertainty of reactor neutrino fluxes is systematically underestimated.

Our analysis will yield its own estimate of the expected uncertainty in $\theta_{12}$.  While this estimate is necessarily quite precise, it will not be accurate.  An accurate determination would require the full covariance matrix of uncertainties for the spectrum generated by each isotope 10 years from now, when the data from these experiments is analyzed.  However such a covariance matrix or isotope by isotope analysis is not available even now.  %In the absence of a covariance matrix, a rough estimate of the uncertainty can be obtained by assuming a model for the systematic bias of the reactor spectrum.  This will be our approach.  As the shape of the systematic bias is not and cannot be known, the uncertainty that we will find will only be reliable at roughly the 50\% level.  We will demonstrate this point by considering two generic models of the bias which indeed lead to different uncertainties, although it will be clear in both cases that the uncertainty in the reactor spectrum is the leading source of uncertainty in $\theta_{12}$.

Our motivation for writing this paper now, when the covariance matrix for the uncertainties is not yet available, is as follows.  In a companion paper \cite{noitracking} we consider the tracking requirements for cosmogenic muons for such experiments. For this, we need to know not the absolute value of the uncertainty in $\theta_{12}$, but rather its expected dependence on the background rejection efficiency.  While the absolute value of the uncertainty that we will obtain is quite approximate,  the current paper nonetheless demonstrates that the uncertainty in $\theta_{12}$ receives a large contribution from systematic errors.  This means that little is lost by increasing the statistical fluctuations via a veto strategy with a large dead time.  In Ref.~\cite{noitracking} we demonstrate that, as a consequence, a very high spallation background rejection efficiency is optimal for the $\theta_{12}$ measurement, higher than that for the mass hierarchy.  This result is quite robust.

\section{The Theoretical Uncertainty has been Underestimated} \label{sottosez}

In this subsection we will motivate our new analysis of the precision of a measurement of $\theta_{12}$ by showing that the uncertainty in the theoretical spectrum \cite{huber}, which has been used in previous determinations of the precision, was greatly underestimated.  Our new study, which will be the subject of Sec.~\ref{mainsez}, will therefore provide a somewhat more reliable determination of this precision.% but also, as will be shown in Ref.~\cite{noitracking}, we will see that larger systematic errors mean that statistical fluctuations will be less important in the determination of $\theta_{12}$ and so a larger cosmogenic background rejection efficiency will be optimal.

Recently a 5 MeV bump in the ratio of the measured reactor neutrino spectrum to the theoretical spectrum of \cite{huber} has been observed by RENO \cite{renobump,renobump2}, Double Chooz \cite{doublebump} and Daya Bay \cite{dayabump}.  The amplitude of this bump is more than 10\%, corresponding to 4$\sigma$ in terms of the theoretical reactor flux uncertainties of Ref.~\cite{huber}.  Therefore it is clear that the difference between the true reactor spectrum and that of Ref.~\cite{huber} is appreciably larger than the subset of the uncertainties which were quantified in that work.  

%Clearly, now that the reactor spectrum has been measured, the determination of $\theta_{12}$ at JUNO and RENO 50 will use the measured spectrum and not the theoretical spectrum. Nonetheless, 

To reassess the validity of a determination of the precision of a measurement of $\theta_{12}$ based on the theoretical spectrum, we will now answer the following question:  What effect does the bump have on a determination of $\theta_{12}$?  

Let us fix the neutrino mass splittings to be
\beq
\m31=2.4\times 10^{-3}{\rm{eV}}^2\hsp
\m21=7.5\times 10^{-5}{\rm{eV}}^2
\eeq
with the normal mass hierarchy and the relevant neutrino mixing angles to be
\beq
\spp2213=0.089,\ 
\spp2212=0.857.
\eeq
We normalize the $\overline{\nu}_e$ flux at JUNO by setting  the number of IBD events to be $10^5$ for a 6 year run at a baseline of 58 km, but we adapt the correct baselines from Ref.~\cite{juno}. 

For the calculation of $\chi^2$, in addition to $\theta_{12}$, we minimize three pull parameters corresponding to the flux normalization of the spectrum and background, with uncertainties of 5\% and 1\% respectively, and also the value $\spp2213$ with an uncertainty of $0.01$.  Variations of the later two uncertainties have little effect on our results.  Then, assuming a perfectly understood nonlinear energy response for the detector and no backgrounds, we find that if the true reactor spectrum is that observed by Daya Bay in Ref.~\cite{dayabump} but it is fit to the theoretical spectrum of Ref.~\cite{huber} then the lowest $\chi^2$ fit would arise with a value of $\spp2212$ which is more than $0.01$ too high.  This is because it leads to less events at the solar oscillation maximum, around 3 MeV, and more events at higher energies, away from the maximum.

By comparison, studies in the literature on the precision of a measurement of $\spp2212$ using the uncertainty reported in Ref.~\cite{huber} estimate a precision of, for example, $0.3\%$ including the uncertainty caused by a model of the detector's nonlinear energy response \cite{whitepaper}.  Thus, were $\theta_{12}$ determined using the theoretical model \cite{huber} of the reactor spectra then the value obtained would differ from the true value by four times the uncertainty reported in, for instance, Ref.~\cite{whitepaper}.  

One might object that it is obvious that, now that the bump has been discovered, one should use the spectrum with the bump for all analyses.  This is of course true.  However it means that a new analysis is needed of the precision with which $\theta_{12}$ can be determined.  This is the goal of the present paper.

\section{The Uncertainty with which $\theta_{12}$ may be Measured} \label{mainsez}

In this note we would like to observe that the precise measurements of the reactor spectrum by the Daya Bay \cite{dayabump} and at the RENO near detector \cite{renobump2} in fact allow for a precise determination of $\theta_{12}$.  An accurate determination of the uncertainty which may be expected in $\theta_{12}$ would require, for each isotope, a covariance matrix of the errors in Refs.~\cite{dayabump}.  Such a set of covariance matrices is not available.  So we simply sum in quadrature the bin per bin statistical and systematic errors reported by Daya Bay and treat them as uncorrelated.

%Of course the uncertainty on $\theta_{12}$ is highly dependent upon the shape of the difference between the true spectrum and measured spectrum.  This shape is not known.  However from Fig.~5 of Ref.~\cite{dayabump} or Fig. 6 of \cite{renobump2} one can estimate the magnitude of the uncertainty and also, due to the binning, one can estimate the variation of the spectrum on energy scales of order the bin size or larger. 

As the entire spectrum, as measured at JUNO or RENO 50, corresponds to only half of a $1-2$ flavor oscillation, only such broad features of the spectrum will be important for measuring $\theta_{12}$.  Therefore even if the underlying reactor spectrum has a rich structure at scales of order 200 keV or smaller, which was not observed in Ref.~\cite{dayabump} due to binning and the finite energy resolution, this will have no effect on the determination of $\theta_{12}$.  On the other hand the determination of the hierarchy depends on $1-3$ oscillations which have a much shorter wavelength and so may be affected by such a substructure in the reactor spectrum \cite{dwyer}, an effect which may even be amplified by the self-calibration of Ref.~\cite{juno}.

The reactor neutrino experiment JUNO will have a very different baseline and total reactor flux from Daya Bay. This leads to different oscillation probabilities.  However, the different oscillation probabilities only affect the normalization of the number of events observed in each bin, and not the fractional uncertainty in the reactor flux.  Therefore, ignoring the somewhat distinct isotope ratios, the fractional uncertainty in the spectrum at each bin at JUNO will be equal to that at Daya Bay.  

To estimate the effect of the unknown spectrum on the determination of $\theta_{12}$, we proceed as follows.  First, we determine the shape of the deformation of the reactor spectrum which would simulate in a shift
\beq
\theta_{12}\rightarrow\tilde{\theta}_{12}=\theta_{12}+\delta\theta_{12} \label{sheq}
\eeq
at JUNO.  With a single detector JUNO can never distinguish such a shift in the reactor spectrum from a shift (\ref{sheq}) in $\theta_{12}$.  We fix the value of $\delta\theta_{12}$ such that such a shift in the reactor spectrum fits Daya Bay's determination of the spectrum with $\chi^2=1$, using the uncertainties reported in Ref.~\cite{dayabump}.  This yields an expected systematic shift in JUNO's measurement of $\spp2212$ of
\beq
\delta(\spp2212)=0.0035.
\eeq
Note that the various degeneracies between the reactor flux uncertainty and uncertainties in the mixing angles, backgrounds, etc. do not affect this calculation, because the $\chi^2$ value of the Asimov data at JUNO with the shifted reactor flux is equal to 0, since the shift in the spectrum has been chosen such that it can be precisely compensated by a shift in $\theta_{12}$.

\begin{figure} %[!tph]
\begin{center}
\includegraphics[width=3in]{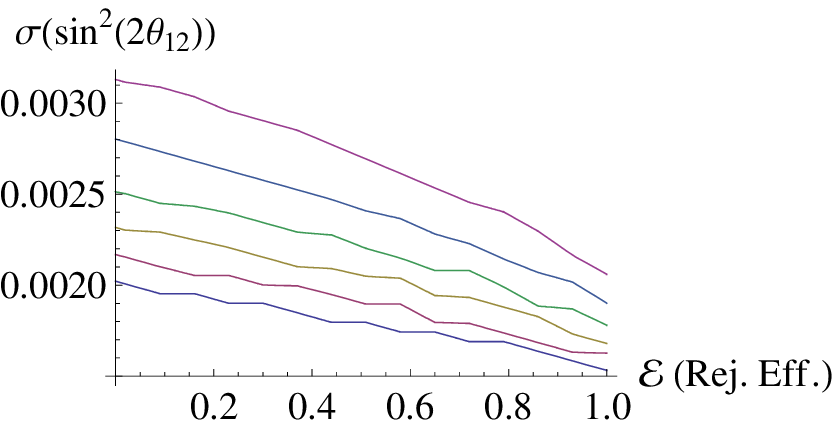}
\includegraphics[width=3in]{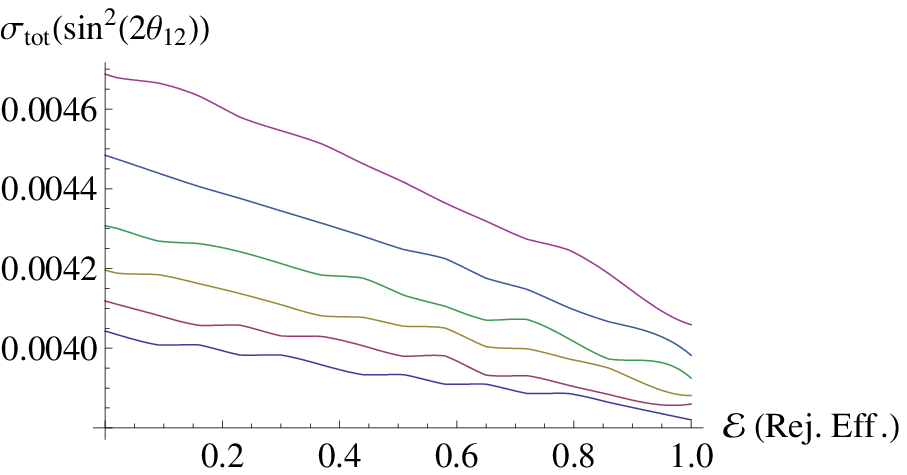}
\caption{The horizontal axis is the ${}^9$Li rejection efficiency.  Each curve represents a different dead time, in ascending order from 0\% to 50\% in steps of 10\%.  {\bf{Left:}} The uncertainty $\sigma$ in the best fit value of $\spp2212$, assuming a perfectly understood reactor spectrum,  optimizing all pull parameters to minimize $\chi^2$.  {\bf{Right:}} The sum in quadrature $\sigma_{tot}$ of the uncertainty $\sigma$ and the shift $\delta(\spp2212)$.}
\label{sigfig}
\end{center}
\end{figure}

The uncertainty in the reactor flux is not responsible for all of the expected uncertainty in $\theta_{12}$.  To determine other contributions to the precision of a measurement of $\spp2212$, we fix the reactor flux to the model of Ref.~\cite{huber} and use the Asimov data set to determine the value of $\spp2212$ for which, when choosing the pull parameters of Sec.~\ref{sottosez} to minimize $\chi^2$, one obtains $\chi^2=1$ after 6 years.  The only background that we consider is cosmogenic ${}^9$Li with the rate given in Ref.~\cite{noimuoni}.  We assume that this background can be rejected with an efficiency $\mathcal{E}$, yielding a fractional dead time.  Various fractional dead times are considered.  We assume that the nonlinear energy response of the detector is perfectly understood, although in practice the uncertainty in the nonlinear energy response will yield a significant contribution to the uncertainty in $\theta_{12}$ \cite{whitepaper}.

This procedure yields the uncertainty in $\theta_{12}$ not including the contribution from the uncertain reactor flux.  The resulting 1$\sigma$ uncertainties are summarized in the left panel of Fig.~\ref{sigfig}.  In the right panel we add the result in quadrature to $\delta(\spp2212)$ to obtain the final uncertainty $\sigma_{\rm{tot}}(\spp2212)$.  As can be seen, using the recent measurements \cite{renobump2,dayabump} one can reduce the uncertainty in $\spp2212$ to about $0.5\%$, which is roughly in line with the stated goals of the experimental collaboration.  A more precise measurement of the reactor spectrum in the future may reduce this \cite{dwyer}, but not beyond the uncertainty displayed in the left panel of Fig.~\ref{sigfig}.

To determine the precision of a measurement of $\spp2212$ if the third and fourth Taishan reactors are not built is straightforward.  These account for 26\% of the total thermal power expected at the Taishan and Yangjiang reactor complexes.  Therefore one can read the resulting uncertainties off of Fig.~\ref{sigfig} by replacing the dead time $\tau$ by 
\beq
\tau^\prime=0.76\tau+0.24.%\hsp \varepsilon^\prime=1.35\varepsilon-0.35.
\eeq

\section{Remarks}

At first glance the fact that our final precision is of the same order as that obtained in previous studies might suggest that this analysis has been trivial.  However we would like to point out that this coincidence is accidental, caused by the fact that the theoretical uncertainties of Ref.~\cite{huber} are similar in magnitude to last year's observational uncertainty \cite{renobump2,dayabump}.  Had we used older data, or last year's data from Double Chooz  \cite{doublebump} then the new uncertainty would have been much larger.  Indeed the two analyses are quite different.  Traditional estimates of the precision of a measurement of $\theta_{12}$ , such as that in Ref.~\cite{whitepaper},  are quite precise as they use the uncertainty in \cite{huber} for which the full covariance matrix is given.  However, for an analysis using the theoretical spectrum of Ref.~\cite{huber}, they are nonetheless inaccurate as that uncertainty was always intended as a lower bound and is now known to be smaller than the true uncertainty by a factor of four.  On the other hand, as the uncertainties in our analysis are observational, there is no such bias.  Nonetheless, as we do not have a covariance matrix for these uncertainties, we assumed that the uncertainties are uncorrelated and thus our estimated uncertainty JUNO's measurement of $\theta_{12}$ is lower than may be expected were JUNO to run today.  On the other hand, Daya Bay and RENO continue to improve the precision of their measurements of the reactor flux, which will reduce the uncertainty in $\theta_{12}$ which will be attained by JUNO, but not beyond that reported in the left panel of Fig.~\ref{sigfig}.

\section* {Acknowledgement}
\noindent
%We are pleased to thank Xin Qian for suggestions and correspondence.  
JE and EC are supported by NSFC grant 11375201.   XZ is supported in part by  NSFC grants 11121092, 11033005 and 11375202.   EC  is also supported by the Chinese Academy of Sciences President's International Fellowship Initiative grant 2015PM063.  MG  is supported by the Chinese Academy of Sciences President's International Fellowship Initiative grant 2015PM007.

%%%%%%%%%%%%%%%%%%%%%%%%%%%%%%%%


\begin{thebibliography}{99}%\setlength{\itemsep}{-2.3mm}

%%%%%%%%%%%%%%%%%%%%%%%%%%%%%%%%%


\bibitem{juno}
  Y.-F.~Li, J.~Cao, Y.~Wang and L.~Zhan,
  ``Unambiguous Determination of the Neutrino Mass Hierarchy Using Reactor Neutrinos,''
  Phys.\ Rev.\ D {\bf 88} (2013) 013008
  [arXiv:1303.6733 [hep-ex]].

\bibitem{reno50}
  S.~B.~Kim,
  ``New results from RENO and prospects with RENO-50,''
  arXiv:1412.2199 [hep-ex].

\bibitem{petcovidea}
  S.~T.~Petcov and M.~Piai,
  ``The LMA MSW solution of the solar neutrino problem, inverted neutrino mass hierarchy and reactor neutrino experiments,''
  Phys.\ Lett.\ B {\bf 533} (2002) 94
  [hep-ph/0112074]. S.~Choubey, S.~T.~Petcov and M.~Piai,
  ``Precision neutrino oscillation physics with an intermediate baseline reactor neutrino experiment,''
  Phys.\ Rev.\ D {\bf 68} (2003) 113006
  [hep-ph/0306017].

\bibitem{parkedifficile}
  S.~J.~Parke, H.~Minakata, H.~Nunokawa and R.~Z.~Funchal,
  ``Mass Hierarchy via Mossbauer and Reactor Neutrinos,''
  Nucl.\ Phys.\ Proc.\ Suppl.\  {\bf 188} (2009) 115
  [arXiv:0812.1879 [hep-ph]].

\bibitem{qianprimo}
  X.~Qian, D.~A.~Dwyer, R.~D.~McKeown, P.~Vogel, W.~Wang and C.~Zhang,
  ``Mass Hierarchy Resolution in Reactor Anti-neutrino Experiments: Parameter Degeneracies and Detector Energy Response,''
  Phys.\ Rev.\ D {\bf 87} (2013) 3,  033005
  [arXiv:1208.1551 [physics.ins-det]].

\bibitem{noisim}
  E.~Ciuffoli, J.~Evslin and X.~Zhang,
  ``Mass Hierarchy Determination Using Neutrinos from Multiple Reactors,''
  JHEP {\bf 1212} (2012) 004
  [arXiv:1209.2227 [hep-ph]].

\bibitem{idea12}
 A.~Bandyopadhyay, S.~Choubey and S.~Goswami,
  ``Exploring the sensitivity of current and future experiments to theta(solar),''
  Phys.\ Rev.\ D {\bf 67} (2003) 113011
  [hep-ph/0302243].


\bibitem{huber}
  P.~Huber,
  ``On the determination of anti-neutrino spectra from nuclear reactors,''
  Phys.\ Rev.\ C {\bf 84} (2011) 024617
   [Erratum-ibid.\ C {\bf 85} (2012) 029901]
  [arXiv:1106.0687 [hep-ph]].

\bibitem{hubertalk}
 P.~Huber,
``Sterile Neutrinos,"
talk given at {\it nuFACT 2012} at the College of William and Mary, Virginia, USA on July 25, 2012

\bibitem{mention}
  G.~Mention, M.~Fechner, T.~Lasserre, T.~A.~Mueller, D.~Lhuillier, M.~Cribier and A.~Letourneau,
  ``The Reactor Antineutrino Anomaly,''
  Phys.\ Rev.\ D {\bf 83} (2011) 073006
  [arXiv:1101.2755 [hep-ex]].

\bibitem{noitracking}
 M.~Grassi, J.~Evslin, E.~Ciuffoli and X.~Zhang,
  ``Vetoing Cosmogenic Muons in A Large Liquid Scintillator,''
  arXiv:1505.05609 [physics.ins-det].

\bibitem{renobump}
 S.~B.~Kim, 
``Observation of Reactor Antineutrino Disappearance at RENO,"
talk given on June 4, 2012 at {\it Neutrino 2012} in Kyoto

\bibitem{renobump2}
  S.~H.~Seo [RENO Collaboration],
  ``New Results from RENO and The 5 MeV Excess,''
  arXiv:1410.7987 [hep-ex].

\bibitem{doublebump}
A. Cabrera,
``Double Chooz III: First results ,"
%H. de Kerret,
%``Results from Double Chooz,"
talk given on May 22, 2014.  % at {\it Neutrino 2014} in Boston
Available at https://indico.lal.in2p3.fr/event/2454/ .

\bibitem{dayabump}
  D.~V.~Naumov [Daya Bay Collaboration],
  ``Recent results from Daya Bay experiment,''
  arXiv:1412.7806 [hep-ex].

\bibitem{whitepaper}
  A.~B.~Balantekin, H.~Band, R.~Betts, J.~J.~Cherwinka, J.~A.~Detwiler, S.~Dye, K.~M.~Heeger and R.~Johnson {\it et al.},
  ``Neutrino mass hierarchy determination and other physics potential of medium-baseline reactor neutrino oscillation experiments,''
  arXiv:1307.7419 [hep-ex].

\bibitem{dwyer}
  D.~A.~Dwyer and T.~J.~Langford,
  ``Spectral Structure of Electron Antineutrinos from Nuclear Reactors,''
  Phys.\ Rev.\ Lett.\  {\bf 114} (2015) 1,  012502
  [arXiv:1407.1281 [nucl-ex]].


\bibitem{noimuoni}
  M.~Grassi, J.~Evslin, E.~Ciuffoli and X.~Zhang,
  %``Showering Cosmogenic Muons in A Large Liquid Scintillator,''
  JHEP {\bf 1409} (2014) 049
  [arXiv:1401.7796 [physics.ins-det]].

\end{thebibliography}
\end{document}